# CoB/Ni-Based Multilayer Nanowire with High-Speed Domain Wall Motion under Low Current Control

Duc-The Ngo[*], Norihito Watanabe, and Hiroyuki Awano[†]

Information Storage Materials Laboratory, Toyota Technological Institute, Nagoya 468-8511, Japan

The spin-transfer torque motion of magnetic DWs in a CoB/Ni-based nanowire driven by a low current density of $(1.12\pm0.8)\times10^{11}$ A m$^{-2}$ has been observed indirectly by magnetotransport measurements. A high DW velocity of 85±4 m/s at zero field was measured at the threshold current density. Upon increasing the current density to $2.6\times10^{11}$ A m$^{-2}$, the DW velocity increases to 197±16 m/s before decreasing quickly in the high-current-density regime owing to nonadiabatic spin-transfer torque at a low damping factor and weak pinning. The addition of B atoms to the Co layers decreased the level of saturation magnetization, Gilbert damping factor, and density of pinning sites, making the CoB/Ni multilayer nanowire favorable for practical applications.

---

[*] Present address: Department of Electrical and Computer Engineering, National University of Singapore, 4 Engineering Drive 3, Singapore 117576.

[†] E-mail address: awano@toyota-ti.ac.jp.

# 1. Introduction

When a spin-polarized electron current hits a magnetic moment, it exerts a torque on the moment, transfers its angular momentum to the moment, and thereby affects the precession motion and switching of the moment. This phenomenon was theoretically predicted by Berger [1] and Slonczewski [2], and subsequently was named the spin-transfer torque (STT). The motion of magnetic domain walls (DWs) caused by an electrical current in magnetic nanostructures is also a consequence of the SST in which the spin-polarized current switches the magnetic moments in the wall. This is nowadays widely applied in spintronic technology such as the DW logic gate [3,4] and racetrack memory [5, 6]. Over the last 15 years, most studies have focused on a NiFe patterned film, a typical soft magnetic material with in-plane magnetic anisotropy and nearly zero magnetocrystalline anisotropy, as it is cheap and highly stable and it is easy to fabricate and control its composition and properties. The motion of DWs with a very high velocity, up to ~200 m/s, has been demonstrated in a number of NiFe-based nanowire devices [3-6]. However, the motion in such in-plane anisotropy films was controllable only at a relatively high current density (~$10^{12}$ A m$^{-2}$) owing to a wide DW and a low spin-torque efficiency (it should be noted that the threshold current density is expected to be proportional to the wall width [7]). A high control current consumes much energy and the heat released from electrical current would sometimes degrade the performance of such devices. Therefore, deceasing of current density is one of the most important technical issues at the moment.

Perpendicularly magnetized thin films have recently been proposed to replace the in-plane NiFe film [8,9] to realize this goal. In the perpendicular magnetic anisotropy films, formation of Bloch-type walls that are 1-2 orders of magnitudes thinner than Néel-type walls in the in-plane films and a high spin-torque efficiency would lower the intrinsic current density by one or even two orders of magnitude. Many authors [10-13] have presented the decrease in threshold current to $(2-5) \times 10^{11}$ A m$^{-2}$ using multilayer nanowires, e.g., Co/Pt, CoFe/Pt, and Co/Ni in which the perpendicular magnetic anisotropy was one of the keys to decrease the critical current density. Nonetheless, layer thickness in those multilayers was normally ~3-20 Å and might be badly influenced by the heat from electrical current. The perpendicular anisotropy is logically threatened to disappear owing to the diffusion of the layers under Joule heating of electrical current.

Among the researchers, Yamanouchi et al. [14] were successful in establishing the motion of magnetic DWs in a perpendicularly magnetized ferromagnetic semiconductor (Ga,Mn)As with a very low current density of about $10^9$ A m$^{-2}$. However, this material (and most ferromagnetic semiconductors) has a Curie temperature far below room temperature and therefore is not realistic for room-temperature devices.

In this article, we present the enhancement of the motion of the magnetic DWs in the CoB/Ni multilayer nanowire at a low current density. The addition of B atoms to the Co layers decreased the density of the pinning sites in the film, enhanced the DW motion, and improved the stability of the multilayer by preventing the diffusion between the Co/Ni interfaces.

## 2. Experimental Methods

A multilayer film of Pt 5 nm/[CoB 0.6 nm/Ni 1.1 nm]4/CoB 0.6 nm/Pt 1 nm was fabricated by radio-frequency (RF) magnetron sputtering using Ar gas. The base vacuum of the deposition chamber was $3\times10^{-8}$ Torr whereas the Ar pressure was maintained at 5 mTorr during the deposition process. The composition of the CoB target was chosen as $Co_{80}B_{20}$ (at.%). The film was grown on a naturally oxidized Si substrate. A nanowire with 300 nm width and 150 μm length was subsequently patterned by electron beam lithography and ion beam etching (Fig. 1). The nanowire was modified to have a planar Hall shape for magnetotransport measurements. A square pad was made at one end of the wire as a source for DW nucleation [15], and the shape of the other end of the wire was modified to be triangular to prohibit the propagation of DW [16]. A Ti/Au electrode pattern produced by photolithography was mounted to the wire for magnetotransport measurements. The magnetic properties of the film specimen were measured using an alternating gradient magnetometer (AGM).

The magnetic DWs were nucleated in the square pad by an Oersted field generated from the 30 ns width, 6.5 MHz pulse current flowing in the Ti/Au electrode deposited on the pad (Fig. 1, electrodes A-B). The motion of DWs in the nanowire was then driven by a DC current ($J_{DC}$, electrodes G-B in Fig. 1). Anomalous Hall effect [17,18] measurement (either electrodes C-D or E-F) was carried out to detect the propagation of DW in the wire. Hysteresis loop measurement on the continuous film specimen (data not shown) using the AGM confirmed that the film exhibited a strong perpendicular magnetic anisotropy with a saturation magnetization of $M_s=5.6\times10^5$ A/m (about 15% lower than that of a Co/Ni-based film [13]) and a uniaxial anisotropy constant of $K_u=3.57\times10^5$ J/m$^3$ (~8% higher than that of a Co/Ni film [12, 13]).

## 3. Results and Discussion

The time-resolved Hall effect voltage signal obtained from the nanowire at a driven current (DC current) of 0.59 mA corresponding to a current density of $j_{DC}=1.12\times10^{11}$ A m$^{-2}$ and an external field of +5 mT is illustrated in Fig. 2(a) and represents the movement of DW along the wire [17,18]. Initially, the wire was magnetically saturated, then magnetization reversal was induced by the Oersted field generated from the pulse current with nucleation of a tiny domain at the region between the wire and the square pad. The DC current sequentially forced the domain (with two walls on two sides) moving along the wire, toward the Hall bar. When the domain moved into the Hall bar, the presence of the domain in the cross bar induced a change in the Hall voltage signal, as seen in Fig. 2(a). The Hall signal switched to a low value when the domain (with the two walls) had passed the Hall bar. The progress of the Hall signal could be interpreted approximately on the basis of a simple schematic shown in Fig. 2(b). Because of a periodic pulse, the domains were nucleated and driven to the wire periodically and the Hall voltage signal appeared to be a periodic pulse. This result looks similar to the DW motion observed previously [17,18]. A square Hall-voltage hysteresis loop [inset of Fig. 2(a)] exhibits a sharp change in the magnetization, proving a fast propagation of a domain through the Hall bar. The square aspect of the hysteresis loop indicated that a reversal occurred through DW nucleation followed by easy DW propagation. It should be noted that the Oersted field released from the driven current was estimated to be about 30 mT, which was much smaller than the coercive field of the sample (see the Hall effect hysteresis loop in the inset of Fig. 2). Therefore, the influence of the Oersted field on the motion of the wall along the wire (described in Fig. 2) could be minor, whereas the effect of the spin-polarized current is essentially considered.

The current dependence of the variation of normalized Hall resistance, $\Delta R_{Hall}$, is shown in Fig. 3. The normalized Hall resistance here was defined as the change in the Hall voltage signal when the domain propagated through the Hall bar [Fig. 2(b)]. Therefore, the normalized Hall resistance became high (1) above a threshold current density of $1.12 \times 10^{11}$ A m$^{-2}$, whereas this value was low (0) below the threshold current density. This indicates that the motion of the magnetic DWs, denoted by a change in Hall resistance, could be induced when the density of the spin-polarized current is above $1.12 \times 10^{11}$ A m$^{-2}$, confirming that it is possible to drive the DW motion in the 300 nm width CoB/Ni nanowire with a threshold current density of $1.12 \times 10^{11}$ A m$^{-2}$ by the STT mechanism. It is important that the threshold current density obtained here was reasonably lower than either $\sim(1-3) \times 10^{12}$ A m$^{-2}$ in the NiFe-based devices [6,7] or $\sim(2-5) \times 10^{11}$ A m$^{-2}$ in a similar multilayer Co/Ni wire [12,13], or lower than the current density in a spin-valve nanowire [19] reported recently. Moreover, it is shown in Fig. 2 that the pulse like signal of Hall voltage is periodic and coherent with the nucleation pulse, presuming the continuous propagation of a multidomain similar to a shift register writing process.

From the pulse like Hall signal, the velocity of DW moving in the Hall bar could be derived [13] from Fig. 2(b): $T_1$ was the time when the front-edge wall of the domain started coming to the Hall bar and $T_2$ was the time that it passed the Hall bar (in L = 500 nm). Therefore, the velocity of the front-edge wall could be referred as L/Δt$_1$ (Δt$_1$ = T$_2$-T$_1$). On the other hand, the velocity of the rear-edge wall could be attained from the time interval Δt$_2$ = T$_4$ − T$_3$. On the other hand, from the phase delay between the signals at the C-D and E-F Hall bars that reflected the time of flight of the wall between two Hall bars, the velocity of the wall in the straight wire (from C to E) was determined. Figure 4 shows the wall velocity in the straight wire area as a function of external magnetic field measured at the threshold current density. The field dependence here is consistent with the following expression [20]:

$$v(H) = \mu_H(H - H_0) + v(J), \qquad (1)$$

where μ$_H$(J) is the DW mobility, J is the current density, and H$_0$ is the "dynamic coercive force". The term v(J) - μ$_H$H$_0$ can be referred as the velocity at zero field.

Using this linear dependence, a zero-field wall velocity of 86±5 m/s was calculated at the critical current density (1.12×10$^{11}$ A m$^{-2}$) with a mobility of 2640±170 (m s$^{-1}$ T$^{-1}$). This matches well with the velocity measured directly at H = 0 (85±4 m/s). It is interesting to note that the field-free wall velocity here was much higher than that of the Co/Ni wire [12,13] or TbFeCo nanowire [18,21]. Therefore, this aspect is very promising for high-speed devices. As DW moved in the region of the Hall bars, the wall velocity (as defined above) was found to be slightly lower than that in the straight wire area but only in the error scale of the measurement. The non-zero DW velocity and linear dependence of the wall velocity on the external field can be attributed to the motion driven by the nonadiabatic torque [20].

The velocities of the front-edge and rear-edge walls were perfectly identical to each other and remained invariable at positions of two Hall bars. These suggest that i) the effect of the pinning on the motion of the walls along the wire was predominantly governed by the material rather than the geometry of the Hall bars and ii) no distortion of the domain and the wall geometry as the domain length was conserved when they were located in the Hall bars. Usually, the distortion of the domain in the Hall bar, denoted by the small difference between the velocities of the front-edge wall (faster) and the rear-edge wall (slower), was only observed at a high applied field (over 90 mT), and can be imagined similarly to the distortion of a balloon, as reported elsewhere [22].

Regarding other interesting points, the time interval $T_3 - T_1$ [see Fig. 2(b)] expresses the period necessary for the whole domain to reach the rear side of the Hall bar, allowing domain size to be estimated. Using this relationship, the average size of the domain was calculated to be 900±35 nm at the critical current density of $1.12 \times 10^{11}$ A m$^{-2}$ and zero field. Under an external field, the domain size was slightly reduced to 630±25 nm at the field of 40 mT, which was similar to theoretical prediction [23]. It is supposed that the external field in this case acted oppositely to the nucleated field from the pulse current, and compressed the domain when it was nucleated. It should be noted that the domain length was conserved when the domain was located in two Hall bars.

The dependence of wall velocity at zero field on controlled current density is shown in Fig. 5. The velocity firstly increased with current density from 85 m/s at the threshold current density to a maximum value of 197±16 m/s at a current density of $2.63 \times 10^{11}$ Am$^{-2}$, then markedly dropped at higher current densities. The variation of wall velocity with current density in this case can be explained qualitatively by referring to the model given in refs.#7 and 24. The model proposed by Tatara et al. [24] predicted that the trend of the wall velocity variation (including a linear increase at low current and a decrease with increasing current at high currents) is a consequence of the nonadiabatic torque driving when the damping factor is low and the pinning effect is weak. In the low current regime, DW velocity was linearly dependent on current density, which is in accordance with the zero-field velocity described in eq. (1) and somehow similar to a previous experimental observation [20]. At a high current density (above $2.63\times10^{11}$ A m$^{-2}$), wall velocity appeared to decrease, indicating that the nonadiabatic parameter β was not zero and not equal to the Gilbert damping factor. This led to a deformation of the wall structure above the Walker breakdown current density [20]. This dependence and linear field-velocity function discussed in previous paragraphs indicated that the motion of the walls in our device was mainly governed by the nonadiabatic term.

In an attempt to explain the decrease in the threshold current density, theoretical models [7,24] are employed, in which the threshold (or intrinsic critical) current density could be referred to as follows [7]:

$$J_c \sim \frac{\alpha e M_s^2}{g \mu_B P} \Delta \frac{1}{|\beta - \alpha|} \quad , \qquad (2)$$

where α and β are the Gilbert damping factor and nonadiabatic spin-transfer torque parameter, respectively; $\Delta$ is the DW width; $M_s$ is the saturation magnetization; $P$ is the spin polarization of the material; $g$ is the gyromagnetic ratio, $e$ is the electron charge, and $\mu_B$ is the Bohr magneton.

As discussed in the previous paragraphs, the magnitude of saturation magnetization of the studied CoB/Ni multilayer film was decreased by ~15%, whereas uniaxial anisotropy was slightly enhanced which subsequently led to a thinner DW. Additionally, the substitution of B for Co is expected to decrease the Gilbert damping factor of the film. Hence, such decreases in saturation magnetization magnitude and wall width could result in a decrease in the threshold critical current. Moreover, the addition of B atoms on the other hand weakens the pinning of the DW by decreasing the number of pinning sites [25] as mentioned in previous paragraphs. This effect also enhances the velocity of the walls. It should be noted that the Walker current at which the wall velocity dropped as seen in Fig. 5, is also expected to be a function of the intrinsic parameters of the materials [26,27].

Additionally, the addition of B to Co would make the devices more stable. B atoms with a small atomic radius would locate at the vacancies in the Co lattice and increase the closed-package degree of the lattice, thus preventing the diffusion between the layers and preserving the magnetic properties of the film under heating caused by the electrical current applied. The addition of B atoms to the Co lattice also decreases the difference in the lattice constant between Co/Ni and results in a smooth Co/Ni interface. It should be noted that 20% addition of B to the Co layers ($Co_{80}B_{20}$) only gives rise to ~12% of resistivity in comparison with pure Co layers. From the technical point of view, these benefits would to enhance the working stability of the devices.

## 4. Conclusions

The motion of magnetic DW in the CoB/Ni multilayer nanowire with a very low current density of $(1.12\pm0.8)\times10^{11}$ A m$^{-2}$ and a high DW velocity of 85±4 m/s has been successfully induced. DW velocity can be raised up to 197±16 m/s by increasing the current density to $2.63\times10^{11}$ A m$^{-2}$. The variation of wall velocity was consistent with the nonadiabatic STT mechanism. These advantages were attributed to the presence of CoB layers with a low Gilbert damping factor, a low saturation magnetization, and a low density of pinning sites. The addition of B also helps in preventing the diffusion between Co and Ni layers and enhances the stability of the multilayer structure and the performance of our device. Using a 30 ns pulse as a writing current, the device could perform shift-register writing of a multidomain state in the wire with an average domain size of 900±35 nm (without field) and a minimum size of 630±25 nm (with field).

**Acknowledgments**

This work was completed with the financial support from the Toyota School Foundation. We thank Professor T. Kato and Professor S. Iwata (Nagoya University) for AGM measurements.

**Figure captions**

**Fig. 1**. Electron microscopy image of the CoB/Ni nanowire with Ti/Au electrodes for magnetotransport measurements. The plus sign denotes the direction of applied field.

**Fig. 2**. (a) Time-resolved Hall voltage signal measured at a driven current of 0.59 mA ($j_{DC}$=1.12×10$^{11}$ A.m$^{-2}$) and external field of 5 mT; (b) Interpretation of the Hall-voltage pulse as domain motion progresses in the Hall bar. The inset shows the field dependence of Hall voltage.

**Fig. 3**. Hall resistance changes as a function of driven current density. The inset shows a time-resolved Hall voltage signal measure at the critical current density and zero field.

**Fig. 4**. External field dependence of the velocity of the wall measured at the threshold current density ($j_{DC}$=1.12×10$^{11}$ A.m$^{-2}$).

**Fig. 5**. Variation of wall velocity as a function of driven current density at zero external field.

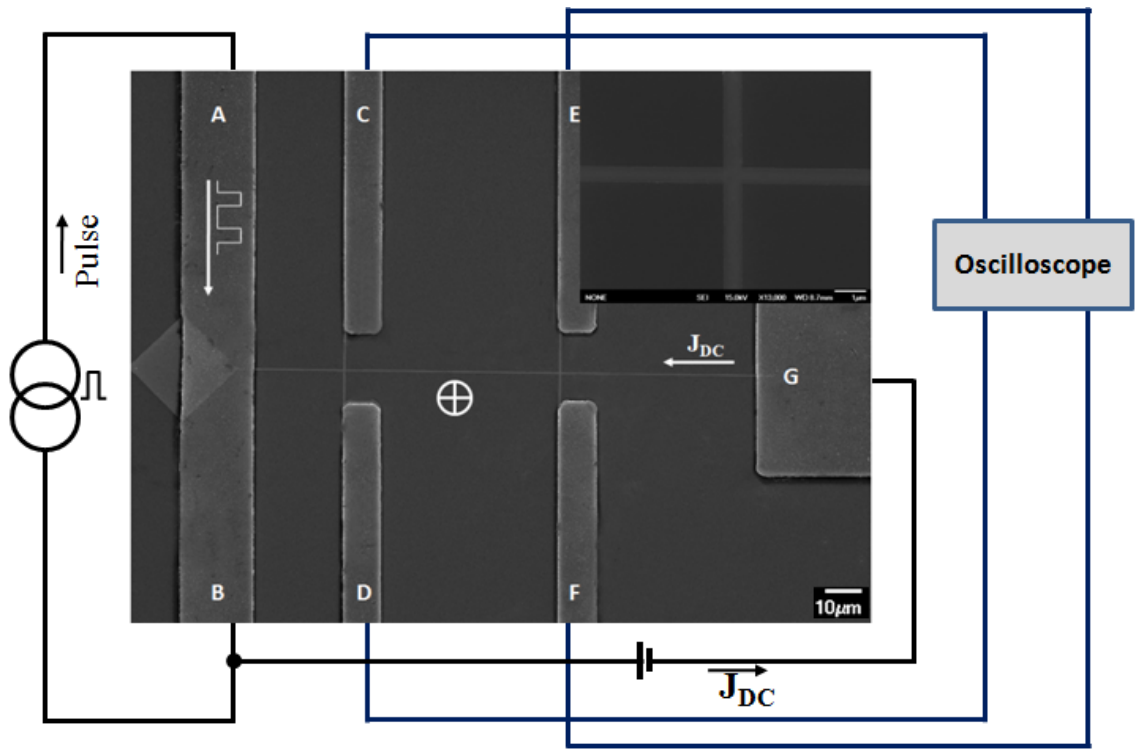

Figure 1

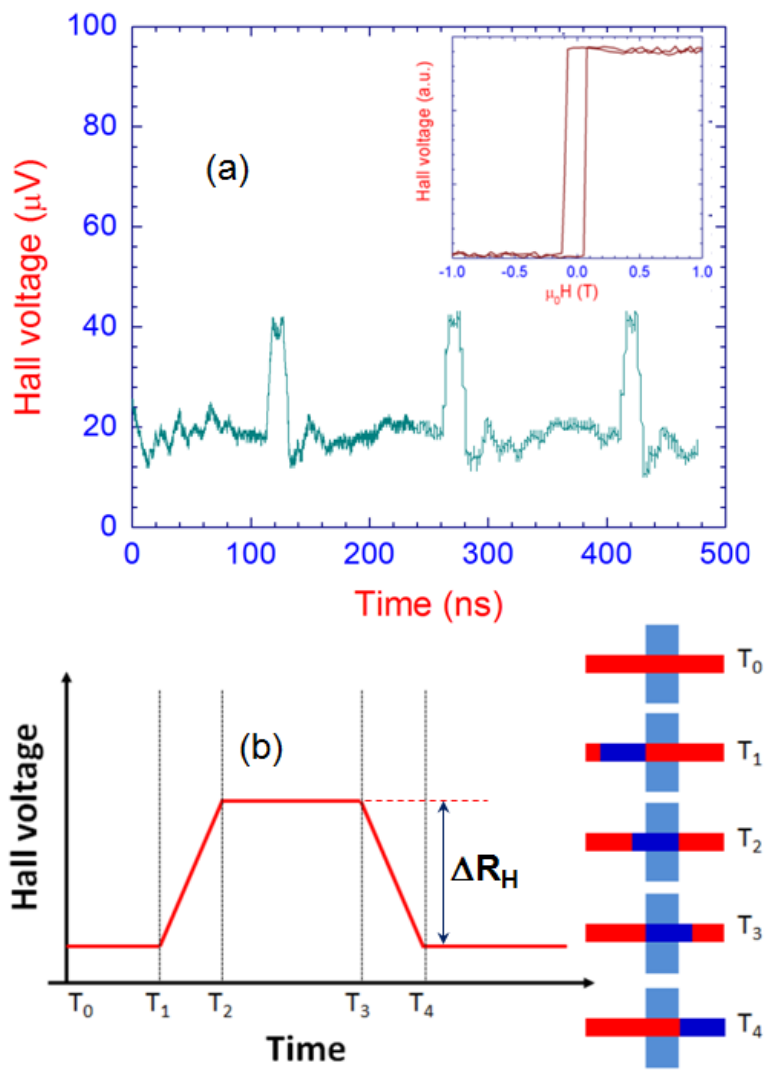

Figure 2

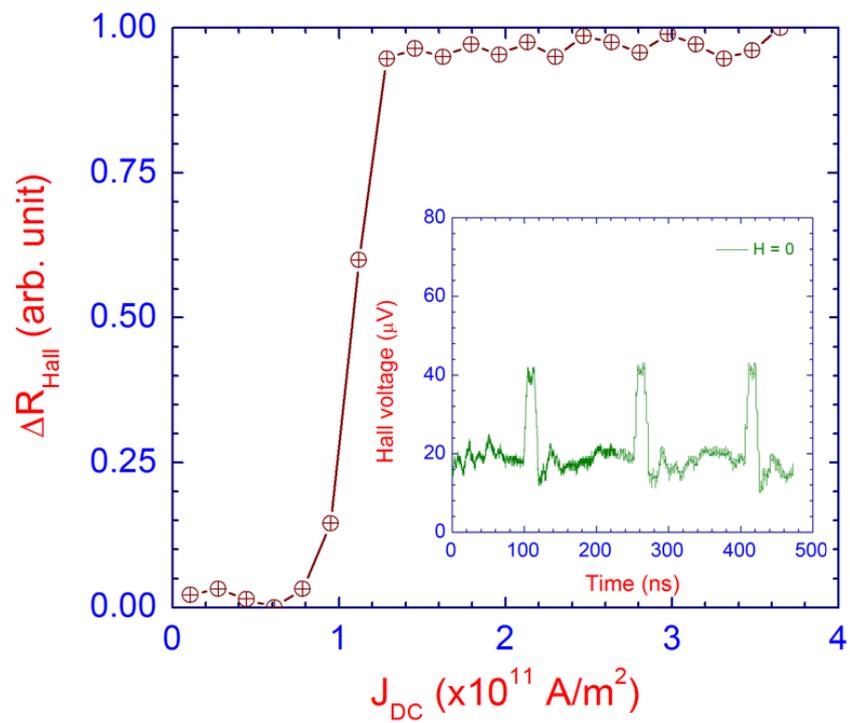

Figure 3

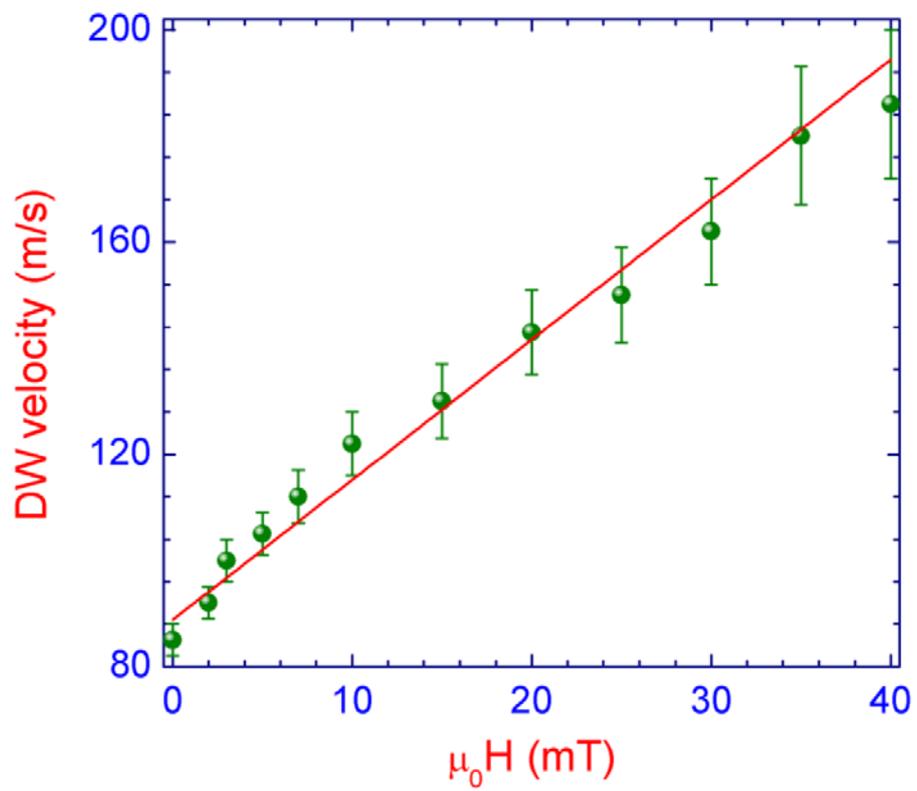

Figure 4

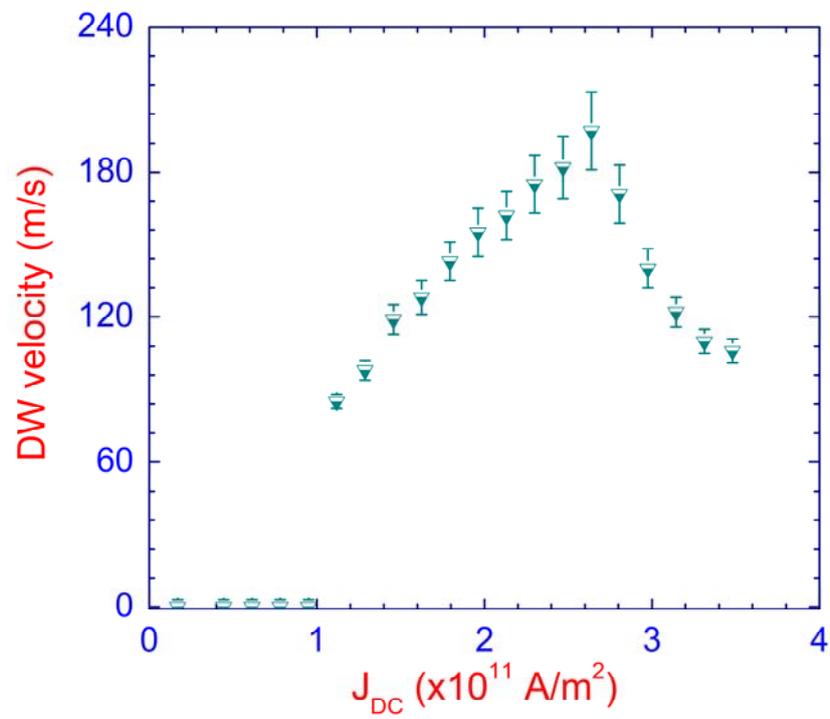

Figure 5